%% file: main.tex
\documentclass{article}
\usepackage[T1]{fontenc}
\usepackage[utf8]{inputenc}

\newif\ifarxiv
\arxivtrue   

\usepackage{ismir} 

\ifarxiv
  \renewcommand{\permission}{}
\fi

\usepackage{amsmath,cite,url}
\usepackage{graphicx}
\usepackage{color}
\usepackage{multirow}
\usepackage{booktabs}
\usepackage{amssymb}
\usepackage{svg}
\usepackage{enumitem}

\title{APEX: Large-scale Multi-task Aesthetic-Informed Popularity Prediction for AI-Generated Music}

\ifarxiv
  \usepackage{authblk}
\fi
\ifarxiv
    \author{
      \parbox{\textwidth}{\centering
        Jaavid Aktar Husain \quad Dorien Herremans \\
         AMAAI Lab, Singapore University of Technology and Design \\
        \small \texttt{jaavidaktar\_husain@mymail.sutd.edu.sg}, \texttt{dorien\_herremans@sutd.edu.sg}
      }
    }
\else
  \twoauthors
    {Jaavid Aktar Husain}
    {AMAAI Lab \\
     Singapore University of Technology and Design \\
     \small \texttt{jaavidaktar\_husain@mymail.sutd.edu.sg}}
    {Dorien Herremans}
    {AMAAI Lab \\
     Singapore University of Technology and Design \\
     \small \texttt{dorien\_herremans@sutd.edu.sg}}
\fi

\def\authorname{J.A. Husain  and D. Herremans}

\begin{document}

\maketitle
\begin{abstract}
    Music popularity prediction has attracted growing research interest, with relevance to artists, platforms, and recommendation systems. However, the explosive rise of AI-generated music platforms has created an entirely new and largely unexplored landscape, where a surge of songs is produced and consumed daily without the traditional markers of artist reputation or label backing. Key, yet unexplored in this pursuit is aesthetic quality. We propose APEX, the first large-scale multi-task learning framework for AI-generated music, trained on over 211k songs (10k hours of audio) from Suno and Udio, that jointly predicts engagement-based popularity signals — streams and likes scores — alongside five perceptual aesthetic quality dimensions from frozen audio embeddings extracted from MERT, a self-supervised music understanding model. Aesthetic quality and popularity capture complementary aspects of music that together prove valuable: in an out-of-distribution evaluation on the Music Arena dataset, comprising pairwise human preference battles across eleven generative music systems unseen during training, including aesthetic features consistently improves preference prediction, demonstrating strong generalisation of the learned representations across generative architectures.
\end{abstract}

\begin{figure}[h!]
    \centering
\includegraphics[width=0.5\textwidth]{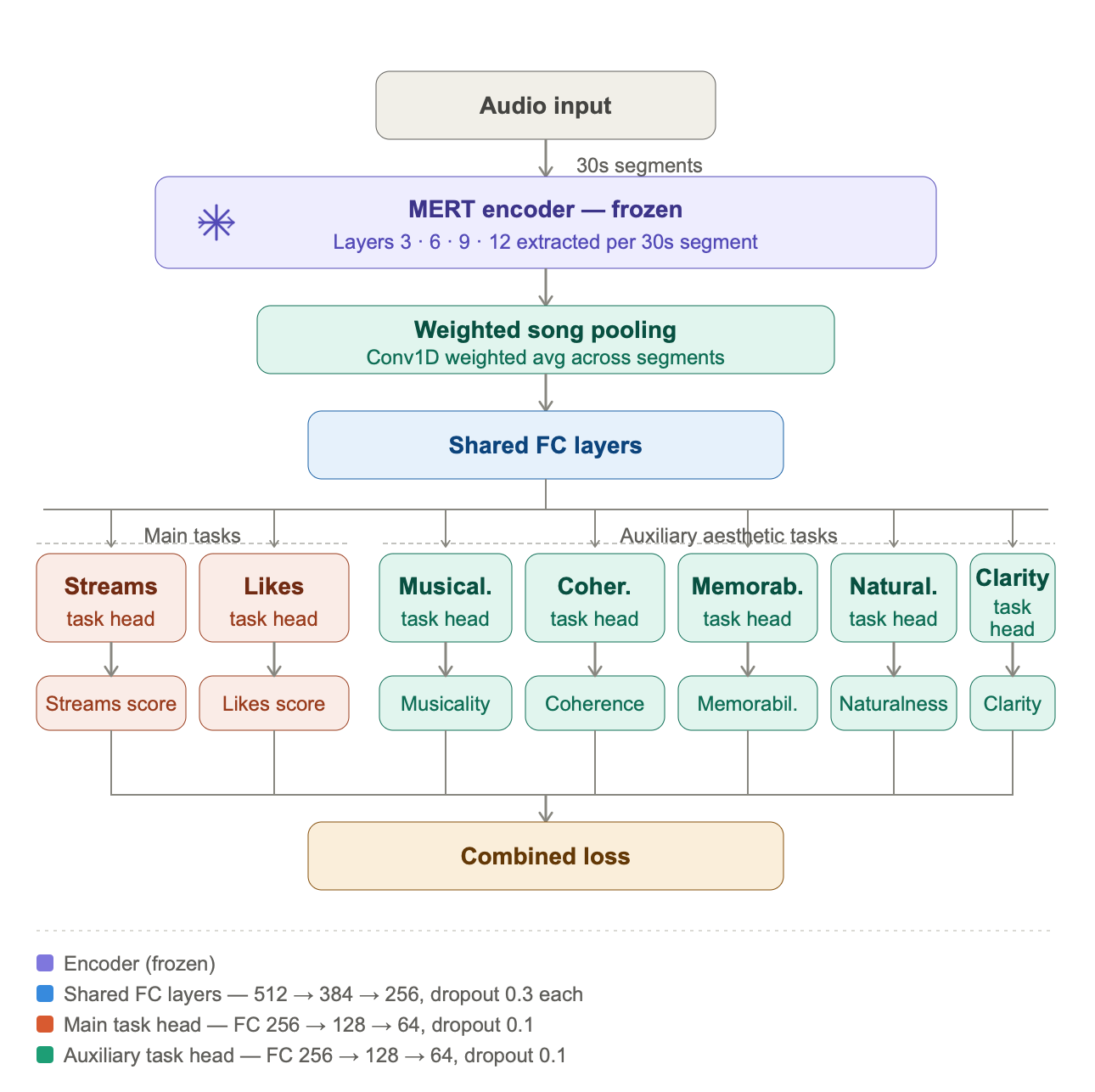}
\caption{Overview of proposed multitask APEX architecture.}
    \label{fig:overview}
\end{figure}
 
\section{Introduction}

Music popularity prediction has been widely studied in the context of commercially released music, where signals such as artist identity, marketing exposure, and historical listener behavior play a central role~\cite{herremans2020hit}. The rapid emergence of AI-generated music platforms has created an entirely new landscape for this problem, where such conventional signals are often absent and models must rely more heavily on the intrinsic properties of the audio. At the same time, research on evaluating the perceptual and aesthetic quality of AI-generated music has grown significantly, with works proposing datasets and metrics capturing dimensions such as coherence, musicality, and audio quality~\cite{yao2025songeval, tjandra2025meta}. However, the relationship between such aesthetic measures and downstream popularity remains largely unexplored.

In this work, we investigate whether aesthetic quality and popularity are intertwined in AI-generated music, and whether modeling them together can yield better popularity predictions. We propose APEX, a multi-task learning framework based on MERT~\cite{li2024mert} audio representations that jointly predicts two engagement-based signals — a streams score and a likes score — alongside five perceptual quality dimensions derived from SongEval~\cite{yao2025songeval}. We train APEX on a large-scale dataset of over 211k AI-generated songs from Udio~\cite{udio2026} and Suno~\cite{suno2026}, and evaluate generalisation on the Music Arena dataset~\cite{kimmusic}, comprising pairwise preference battles between tracks from eleven generative music systems unseen during training.

Our results reveal that aesthetic quality and popularity capture complementary but distinct aspects of AI-generated music, with the full multi-task configuration performing comparably to the popularity-only baseline. Notably, aesthetic dimensions are predicted with considerably higher accuracy, and APEX predictions serve as meaningful proxies for human preference in a fully out-of-distribution setting, demonstrating strong generalisation across unseen generative architectures.

In summary, this work makes the following contributions:
\begin{itemize}
    \item We propose APEX, the first large-scale multi-task framework for jointly predicting popularity and aesthetic quality in AI-generated music, trained on over 211k songs.
    \item We provide an empirical analysis showing that aesthetic quality and popularity capture complementary but distinct signals in AI-generated music, with both dimensions being learnable from audio representations alone.
    \item We conduct a systematic ablation study across 24 experimental conditions examining loss strategy, shared layer depth, input mode, and task configuration.
    \item We demonstrate good out-of-distribution generalisation through a pairwise human preference experiment on the Music Arena dataset, covering eleven unseen generative music systems.
\end{itemize}

\section{Related work}




Music popularity prediction, often termed ``Hit Song Science,'' has evolved significantly since 2008 when it was questioned whether this field could be considered a rigorous science~\cite{pachet2008hit}. Early work focused on extracting acoustic characteristics to predict song success, with studies pioneering dance hit prediction~\cite{herremans2014dance} using supervised learning on audio features. The introduction of deep learning marked a significant shift, with convolutional neural networks learning features directly from mel-spectrograms~\cite{yang2017revisiting}, inspiring numerous studies on datasets from Spotify and other streaming platforms~\cite{middlebrook2019song,kim2021music,sebastian2024beyond}. However, most of these models remain relatively small, relying on handcrafted features due to limited dataset sizes. As the field matured, researchers recognised that audio features alone provided an incomplete picture, leading to multimodal approaches integrating audio, lyrics, and metadata~\cite{martin2020multimodal,zhao2023analysis}. Historical streaming metrics were shown to provide valuable predictive signals~\cite{cabansag2025prediction}, while social media and listening statistics emerged as another important dimension~\cite{herremans2020hit,yee2022predicting,kim2014nowplaying,tsiara2020using,aum2023can,rompolas2024predicting,wu2024leveraging}. Lyrical content has also been explored through semantic analysis~\cite{singhi2015can,dhanaraj2005automatic} and language model embeddings~\cite{choudhary2025lyrics,vavaroutsos2023hsp}, and musical homophily was shown to improve prediction precision through social network influence parameters~\cite{reisz2024quantifying}. Sequential models such as LSTMs proved effective for modelling temporal popularity patterns~\cite{li2021lstm,liu2022music}, while unconventional approaches including neurophysiological methods have also been explored~\cite{merritt2023accurately,arora2024soundtrack}.

Parallel to this, specialised methods for evaluating AI-generated music have emerged~\cite{xiong2023comprehensive,kader2025survey}. SongEval~\cite{yao2025songeval} provides expert aesthetic ratings across multiple dimensions, while AudioBox Aesthetics~\cite{tjandra2025meta} focuses on perceptual quality metrics aligned with human aesthetic judgements. These evaluation methods have become particularly valuable for training generative music models through techniques like Direct Preference Optimization. Metrics such as Fréchet Audio Distance~\cite{kilgour2018fr} and MuQ-Eval~\cite{zhu2026muq} offer automated quality assessment, though comprehensive evaluations reveal that many objective metrics align poorly with human musical preferences~\cite{kader2025survey}. Despite this evolution from audio-only features to sophisticated multimodal approaches, a significant gap remains: virtually no work has addressed predicting the popularity of AI-generated music specifically, highlighting the need for dedicated models in this space.

\section{Proposed APEX model}

The overall architecture of our proposed method is shown in Figure \ref{fig:overview}.

\subsection{MERT Encoder}
We adopt MERT \cite{li2024mert}, a self-supervised transformer encoder for music representation learning. It uses a dual-teacher pretraining framework combining an acoustic teacher based on RVQ-VAE and a musical teacher based on the Constant-Q Transform (CQT), enabling it to capture both low-level acoustic features and higher-level musical structure. This makes MERT well-suited for music popularity prediction, which requires modeling deeper musical characteristics beyond surface-level audio cues. Moreover, our cross-platform experiments (Section~\ref{sec:human_preference}) show that MERT embeddings generalise to unseen generative models, indicating that they capture fundamental musical properties. 

\subsection{Multitask approach}

\subsubsection{Main task: Streams- and likes-score}

To derive a continuous popularity score from raw stream counts, we first map each track's stream count to its percentile rank within the dataset, normalising the distribution across tracks regardless of absolute magnitude. The raw percentile is then transformed via a power function
\begin{equation}
    s = \left(\frac{p}{100}\right)^{\alpha} \times 100,
\end{equation}
where $p$ is the percentile rank and $\alpha = \frac{\ln 0.5}{\ln 0.8} \approx 3.106$. This exponent is chosen such that a track at the 80th percentile receives a score of 50, deliberately compressing the upper tail of the distribution and penalising tracks that are merely popular relative to the dataset but not exceptionally so. The resulting score $s \in [0, 100]$ is right-skewed, rewarding only tracks with strong percentile standing. An identical procedure is applied to derive the likes score, with like counts substituted for stream counts. This type of score ports across datasets and provides a score that other models can use for potential DPO or reinforcement learning.

\subsubsection{Auxiliary tasks: Aesthetics scores}
We incorporate auxiliary tasks that model perceptual attributes of music using SongEval \cite{yao2025songeval}, a benchmark dataset with expert aesthetic ratings for evaluating songs across multiple dimensions. SongEval provides five scores—\emph{coherence, musicality, memorability, clarity, naturalness}—each ranging from 1 to 5, capturing different dimensions of perceived music quality. We use the model trained on SongEval dataset released by the authors\footnote{\url{https://github.com/ASLP-lab/SongEval}} to generate these scores for all songs and use them as labels for the auxiliary tasks. SongEval provides multi-dimensional, human-aligned aesthetic evaluations that complement traditional popularity signals.

\subsubsection{Combining losses}
\label{sec:losses}
Each task head has a loss $\mathcal{L}_i$. To combine them we explore three strategies in Section~\ref{sec:ablatio_experiment}. 

The first strategy uses an \textbf{equal-weight sum}, $\mathcal{L}_{total} = \sum_{i=1}^{T} \mathcal{L}_i$, where $\mathcal{L}_i$ is the MSE loss for task $i$. The second applies \textbf{manual task weighting}, $\mathcal{L}_{total} = \sum_{i=1}^{T} w_i \mathcal{L}_i$, assigning $w_i = 5.0$ to popularity tasks and $w_i = 1.0$ to aesthetic tasks to prioritise the harder primary objectives. The third strategy adopts an \textbf{uncertainty-based learned weighting}~\cite{kendall2018multi}, where each task is assigned a learnable uncertainty parameter $\sigma_i$ that automatically balances task contributions during training:
\begin{equation}
    \mathcal{L}_{total} = \sum_{i=1}^{T} \frac{1}{2\sigma_i^2} \mathcal{L}_i + \log \sigma_i
\end{equation}
This formulation allows the model to automatically balance task contributions based on their homoscedastic uncertainty preventing from destabilizing the shared representation learning.


\section{Experimental setup}

\subsection{Dataset}

We construct our dataset by combining subsets of two large-scale AI-generated music repositories: Udio-126k\footnote{\url{https://huggingface.co/datasets/sleeping-ai/Udio-126K}} and Suno-307k\footnote{\url{https://huggingface.co/datasets/sleeping-ai/suno-307K}}. The music is these repositories is sourced from Udio and Suno respectively. Each of the songs is accompanied by `streams' counts, `likes' counts, and other meta-data. We remove songs with zero streams, any duplicated songs, corrupted audio files, as well as those released within two weeks of the dataset release to avoid recency bias. We retain approximately 124k songs per platform. Since the raw Suno subset is larger, stratified sampling is applied to match the size of the Udio subset while preserving the streams score distribution. The combined $\sim$248k songs are split into train, test and validation sets at 85\%, 10\%, and 5\% respectively using stratified sampling, yielding a training set of $\sim$211k songs corresponding to approximately 10k hours of audio.

\subsection{Embedding extraction}

Audio embeddings are extracted from each song using MERT-v1-95M \cite{li2024mert}. Each audio file is first converted to mono and resampled to 24~kHz to match the model's expected sampling rate. The audio is then segmented into non-overlapping 30-second windows, with shorter final segments zero-padded to maintain a consistent length.

Each segment is passed through MERT, and hidden states are extracted from four intermediate transformer layers (3, 6, 9, and the final layer), selected to provide evenly spaced coverage across the full network depth. This is motivated by the MERT paper\cite{li2024mert}, which shows that earlier layers capture acoustic-level features while deeper layers model higher-level musical abstractions. Multi-layer aggregation of MERT representations has also been adopted in prior work on music understanding\cite{papaioannou2025universal}, supporting the use of representations from multiple layers over a single layer alone.. The hidden states from each layer are mean-pooled across the time dimension to produce a 768-dimensional vector per layer, yielding four vectors of dimension 768 per segment. These are aggregated into a single 768-dimensional embedding using a 1D convolutional layer (Conv1d) with learned weights, which acts as a trainable linear combination across the four layer representations.

\subsection{Training}

All models are trained using the AdamW optimiser with a learning rate of $1 \times 10^{-4}$, weight decay of $1 \times 10^{-4}$, and a cosine annealing learning rate scheduler. Training is performed with a batch size of 512 per GPU across 4 NVIDIA Tesla V100 GPUs using Distributed Data Parallel (DDP). Mixed precision training is applied throughout to improve efficiency. Early stopping is applied based on validation loss.

\subsubsection{Input Modes}
We experiment with two input modes that differ in how song-level representations are constructed from segment embeddings. In \textbf{segment mode}, each 30-second segment is treated as an independent training sample, allowing the model to learn from fine-grained temporal windows of audio directly. In \textbf{song mode}, all segment embeddings for a given song are averaged into a single vector prior to training, providing a holistic song-level representation. At evaluation time, segment-mode models aggregate their per-segment predictions by averaging across all segments of a song before computing metrics.

\subsubsection{Task Configurations}
We experiment with two task configurations. The \textbf{popularity} configuration trains two output branches — one for streams score and one for likes score — focusing exclusively on the engagement-based prediction objectives. The \textbf{full} configuration trains all seven branches jointly, adding five aesthetic quality branches (coherence, musicality, memorability, clarity, and naturalness) alongside the two popularity branches, enabling multi-task learning across both engagement and perceptual quality signals.

\input{tables/table1}

\subsubsection{Model Architecture Variants}
We investigate two shared layer configurations. The first uses \textbf{two shared layers} with dimensions $768 \rightarrow 512 \rightarrow 256$, and the second uses \textbf{three shared layers} with dimensions $768 \rightarrow 512 \rightarrow 384 \rightarrow 256$, adding an intermediate layer to increase representational capacity. In both cases, each shared layer consists of a linear transformation followed by batch normalisation, GELU activation, and dropout with rate 0.3. Each task-specific branch follows the structure $256 \rightarrow 128 \rightarrow 64 \rightarrow 1$ with the same normalisation and activation pattern and a dropout rate of 0.1. Popularity branch outputs are scaled to the range $[0, 100]$ via a sigmoid activation, while aesthetic quality branches are scaled to $[1, 5]$.

\subsubsection{Experimental Grid}
Combining the three loss strategies (Section~\ref{sec:losses}), two shared layer configurations, two input modes, and two task configurations yields a total of $3 \times 2 \times 2 \times 2 = 24$ experimental conditions, which are evaluated in Section~\ref{sec:ablatio_experiment}.





\subsection{Pairwise human preference experiment}
\label{sec:human_preference}

We evaluate whether predicted popularity and aesthetic scores from the APEX model can be used to predict
human pairwise music preference in an out-of-distribution setting. We therefore evaluate on the Music Arena Dataset~\cite{kimmusic}, comprising pairwise 
preference `battles' between tracks generated by eight generative music systems. The models used to generate include state-of-the-art open and commercial models: Sonauto \cite{sonauto2026}, ACEStep \cite{gong2025ace}, ElevenLabs \cite{elevenlabs2026}, MusicGen \cite{copet2023musicgen}, Riffusion \cite{riffusion2025fuzz}, and Lyria \cite{deepmind2025lyria}. We filtered the last 4 months of data from the Music Aren Dataset and kept only battles with valid binary preferences (A or B), removing ties/both-bad votes, as well as battles with missing audio files, resulting in 1,259 battles. Each battle presents two tracks generated from the same prompt, with a human-provided preference label. 
The dataset contains 780 instrumental and 479 vocal tracks.

For each battle, we compute three feature types for each APEX-predicted score $f$. First, difference scores $\Delta_f = f_a - f_b$ capture the absolute advantage of track $A$ over track $B$. Second, ratio scores $r_f = f_a / (f_b + \epsilon)$ capture the \textit{relative} gap, where $\epsilon = 10^{-9}$ prevents division by zero. Third, interaction terms $\Delta_f \times \mathbb{1}_{\text{instrumental}}$ model the hypothesis that aesthetic dimensions contribute differently to preference depending on vocal presence. We additionally include a binary instrumental indicator $\mathbb{1}_{\text{instrumental}}$ as a standalone feature. Applied across 10 APEX score dimensions (predicted streams, likes, coherence, musicality, memorability, clarity, naturalness, combined popularity, combined SongEval, and combined overall), this yields $10 \times 3 + 1 = 31$ features in total.

We train five baseline classifiers using stratified 10-fold cross-validation to preserve class distribution across folds. The models are as follows: (1) Logistic Regression with L2 regularization (C = 0.1, max iterations = 1000) and balanced class weights; (2) Random Forest with 300 estimators, a maximum depth of 4, and balanced class weights; (3) XGBoost with 300 estimators, a learning rate of 0.05, maximum depth of 4, and a positive class weight scaled by the inverse class frequency ratio (674/585) to address class imbalance; (4) AdaBoost with 300 estimators and a learning rate of 0.05; and (5) a Support Vector Machine (SVM) with hyperparameters selected via grid search. 
We also compare against a naive rule-based approach that compares which of the two audio files has the highest predicted (sum of) scores of the selected feature set. The audio with the highest total score is selected as the the preferred audio. We note that the dataset set exhibits mild class imbalance (674 vs.\ 585 instances for tracks A and B respectively), which we account for through class-weight rebalancing in all applicable classifiers.

This experiment will test the generalizability of our model as it includes music generated by a large number of state-of-the-art models\footnote{\scriptsize{\texttt{riffusion-fuzz-1-0}, \texttt{riffusion-fuzz-1-1}, \texttt{sonauto-v2-2}, \texttt{magenta-rt-large},\texttt{sonauto-v3-preview}, \texttt{musicgen-small}, \texttt{musicgen-medium}, \texttt{elevenlabs-music-v1}, \texttt{lyria-3-30s}, \texttt{lyria-3-pro-preview}, and \texttt{acestep-1.5-turbo-1.7b}}}.

\section{Results}

\subsection{Ablation study}
\label{sec:ablatio_experiment}
Table \ref{table1} reports the popularity prediction performance across all 24 experimental conditions on the held-out test set (10\% of the full dataset which is around 25k songs). Overall, results are consistent across configurations, with MSE ranging from 699–714 and MAE from 21.0–22.3 for streams score, and MSE from 659–677 and MAE from 19.97–21.68 for likes score. Pearson and Spearman correlations range from 0.33–0.35 and 0.33–0.35 for streams score, and 0.39–0.42 and 0.40–0.42 for likes score respectively across all conditions.

Song mode consistently outperforms segment mode across all loss strategies and layer configurations, yielding lower MSE and MAE alongside higher correlations, suggesting that averaging segment embeddings into a holistic song-level representation is more effective. The three-layer shared architecture yields marginal improvements in MSE over the two-layer variant, indicating that additional representational capacity provides limited benefit beyond a point. Across loss strategies, uncertainty-based weighting (Models C and F) achieves the lowest MSE and MAE and highest correlations, while manual weighting performs comparably to equal weighting. Notably, the full task configuration — jointly predicting popularity and aesthetic quality — performs comparably to the popularity-only baseline across all metrics, suggesting that aesthetic auxiliary tasks capture complementary information without compromising popularity prediction performance. The best overall configuration is Model C (uncertainty loss, two shared layers, song mode, full task), achieving overall lower errors and better co-relations. This is also supported by the aesthetic evaluation (Section \ref{sec:aethetic_evaluation_results}) and the human preference experiment (Section \ref{sec:human_preference_results}).

\subsection{Aesthetic prediction}
\label{sec:aethetic_evaluation_results}
\input{tables/aesthetics}

Table \ref{tab:song_level_metrics} reports aesthetic prediction performance across all models. The models perform well on the task, with MSE ranging from 0.166 to 0.289 and Pearson correlations from 0.59 to 0.75 across all SongEval dimensions and models. Model C achieves the strongest performance overall, with MSE as low as 0.166 for coherence and naturalness, and Pearson correlations of 0.734–0.751 across the five SongEval dimensions, while Model F follows closely with Pearson correlations of 0.687–0.705. Weighted loss configurations (Models B and E) perform the least good, suggesting that up-weighting popularity tasks at the expense of aesthetic tasks degrades aesthetic prediction without meaningfully improving popularity prediction. Naturalness is consistently the best-predicted dimension across all models in terms of both MSE and correlation, while memorability is the most challenging. These results indicate that perceptual aesthetic dimensions are learnable from MERT audio embeddings, even though they do not directly translate into improved popularity prediction. Also we can the the best performing model is aesthetic prediction is also the best in popularity prediction tasks which supports our assert of modelling popularity and aesthetics together.

\input{tables/arena.tex}

\subsection{Pairwise human preference experiment}
\label{sec:human_preference_results}

Table~\ref{tab:arena} reports preference prediction performance for Model C---the best-performing model from our ablation study---overall and stratified by vocal presence.

Even among the naive rule-based baselines, the value of aesthetic features is apparent: the rule using all predicted scores (AUC = 0.535) outperforms using likes alone (AUC = 0.518), suggesting that aesthetic dimensions contribute complementary signal beyond engagement-based features. This is further confirmed by the classifier results, where models with aesthetic features consistently outperform the same models without aesthetic features across all classifiers. The SVM achieves AUC of 0.642 and F1 of 0.595 with aesthetic features compared to AUC of 0.614 without, demonstrating that aesthetic dimensions provide meaningful additional signal for preference prediction.

Performance is consistently higher on instrumental tracks than on vocal tracks across all configurations---for example, the SVM with aesthetic features achieves AUC of 0.686 on instrumental tracks compared to 0.560 on vocal tracks. We attribute this gap to the presence of vocal artefacts in AI-generated singing, which can introduce perceptual inconsistencies that are difficult to capture from audio embeddings alone. Instrumental tracks present a cleaner signal for aesthetic-based preference modelling.

Importantly, all generative systems in the Music Arena dataset were entirely unseen during training. The fact that APEX achieves above-chance preference prediction across these unseen systems demonstrates that the learned representations generalise well beyond the Suno and Udio training distribution, indicating that MERT-based audio embeddings capture fundamental musical properties that transfer across generative architectures. Overall, these results suggest that APEX predictions serve as meaningful proxies for human preference in a fully out-of-distribution setting.

\section{Conclusion}

We presented APEX, the first large-scale multi-task framework for jointly predicting popularity and aesthetic quality in AI-generated music, trained on over 211k songs from Suno and Udio using frozen MERT audio embeddings. Our ablation study across 24 experimental conditions demonstrates that uncertainty-based loss weighting and song-level embedding aggregation yield the best overall performance, and that the full multi-task configuration captures both popularity and aesthetic dimensions without degrading either---positioning APEX as a versatile audio understanding framework beyond popularity prediction alone.
 
Aesthetic quality dimensions are well captured by MERT representations (Pearson up to 0.75), and strikingly, the best-performing model on popularity is also the best on aesthetics. While jointly modelling the two does not improve engagement-based popularity prediction over a popularity-only baseline, the two objectives capture complementary aspects of music that together prove valuable downstream. Our out-of-distribution pairwise human preference experiment on the Music Arena dataset demonstrates that including aesthetic features consistently improves preference prediction across eleven unseen generative music systems, indicating that the learned representations capture fundamental musical properties that transcend specific generative architectures. Future work could explore vocal-aware modelling to close the remaining performance gap on vocal tracks.
 
APEX is released as open source model to support the community in this rapidly growing area.\footnote{Code: \url{https://github.com/AMAAI-Lab/apex }
Model: \url{https://huggingface.co/amaai-lab/apex}}

\section{Acknowledgments}
This work has received funding from grant no. SUTD SKI 2021\_04\_06 and from MOE grant no. MOE-T2EP20124-0014

\section{AI Usage Statement}
We acknowledge the use of chatGPT and Claude for grammar improvements.

\bibliography{amaai,ISMIRtemplate}

\end{document}

%% file: tables/table1.tex
\begin{table*}[htb!]
\footnotesize
\begin{tabular}{@{}lllll|cccc|cccc@{}}
\toprule
 & & &     &        & \multicolumn{4}{c}{\textbf{Streams Score}}   & \multicolumn{4}{c}{\textbf{Likes Score}}     \\ \midrule
 \textbf{Model} & Loss & FC
 &   \textbf{Mode}      &   \textbf{Task}         & \textbf{MSE}     & \textbf{MAE}   & \textbf{Pearson} & \textbf{Spearman} & \textbf{MSE}     & \textbf{MAE}   & \textbf{Pearson} & \textbf{Spearman} \\ \midrule
\multirow{4}{*}{A}     & \multirow{4}{*}{Equal} & \multirow{4}{*}{2}                 & segment              & popularity           & 713.98 & 22.29 & 0.34  & 0.33  & 677.17  & 21.68 & 0.40 & 0.40 \\
& &  & segment & full       & 712.40  & 22.24 & 0.34    & 0.33     & 675.42  & 21.62 & 0.40    & 0.40     \\
& &  & song    & popularity & 702.62  & 21.00 & 0.34    & 0.35     & 662.20  & 20.00 & 0.41    & 0.42     \\
& &  & song    & full       & 702.12  & 21.02 & 0.34    & 0.35     & 662.69  & \textbf{19.97} & 0.41    & 0.42     \\ \midrule
\multirow{4}{*}{B}   & \multirow{4}{*}{Weighted} & \multirow{4}{*}{2}     & segment              & popularity           & 713.03 & 22.25 & 0.34  & 0.33  & 676.59  & 21.65 & 0.40 & 0.39 \\
& &  & segment & full       & 711.11  & 22.18 & 0.34    & 0.34     & 673.97  & 21.53 & 0.40    & 0.40     \\
& &  & song    & popularity & 700.44  & 21.02 & 0.35    & 0.35     & 660.42  & 20.02 & 0.41    & 0.42     \\
& &  & song    & full       & 702.46  & 21.14 & 0.34    & 0.35     & 662.71  & 20.09 & 0.41    & 0.42     \\ \midrule
\multirow{4}{*}{C}  & \multirow{4}{*}{Uncert.} & \multirow{4}{*}{2}                   & segment              & popularity           & 710.95 & 22.17 & 0.34  & 0.33  & 673.81  & 21.53 & 0.40 & 0.40 \\
& &  & segment & full       & 709.44  & 22.15 & 0.34    & 0.34     & 672.60  & 21.53 & 0.41    & 0.40     \\
& &  & song    & popularity & 700.56  & 21.03 & 0.35    & 0.35     & 661.02  & 19.99 & 0.41    & 0.42     \\
& &  & song    & full       & 701.12  & \textbf{20.98} & 0.35    & 0.35     & 661.76  & \textbf{19.97} & 0.41    & 0.42     \\ \midrule
\multirow{4}{*}{D}  & \multirow{4}{*}{Equal} & \multirow{4}{*}{3}                   & segment              & popularity           & 712.66 & 22.24 &  0.34 & 0.33  & 675.15  & 21.61 & 0.40 & 0.40 \\
& &  & segment & full       & 711.69  & 22.21 &   0.34  &  0.34    & 674.16  & 21.56 &  0.40   &   0.40   \\
& &  & song    & popularity &  699.80 & 21.13 &  0.35   &   0.35   &  659.69 & 20.12 &  0.41   &  0.42    \\
& &  & song    & full       &  \textbf{699.76} & 21.20 &  0.35   &  0.35    & \textbf{659.21} & 20.14 &  0.41   &  0.42    \\ \midrule

\multirow{4}{*}{E}  & \multirow{4}{*}{Weighted} & \multirow{4}{*}{3}                   & segment              & popularity           & 711.08 & 22.17 &  0.34 & 0.34  & 673.52  & 21.51 & 0.40 & 0.40 \\
& &  & segment & full       &  711.22 & 22.20 &   0.34  &  0.34    & 673.97  & 21.55 &  0.40   &   0.40   \\
& &  & song    & popularity &  699.84 & 21.15 &  0.35   &   0.35   & 659.62  & 20.11 &  0.41   &   0.42   \\
& &  & song    & full       & 700.00  & 21.19 &   0.35  &   0.35   & 659.62 & 20.16 &  0.41   &  0.42     \\ \midrule

\multirow{4}{*}{F} & \multirow{4}{*}{Uncert.} & \multirow{4}{*}{3}  & segment              & popularity           & 710.01 & 22.15 & 0.34  & 0.34  & 672.50  & 21.51 & 0.41 & 0.40 \\
& &  & segment & full       & 709.14  & 22.12 & 0.34    & 0.34     & 671.21  & 21.46 & 0.41    & 0.40     \\
& &  & song    & popularity & 700.72  & 21.21 & 0.35    & 0.35     & 660.30  & 20.15 & 0.41    & 0.42     \\
& &  & song    & full       & 700.47  & 21.14 & 0.35    & 0.35     & 660.00  & 20.08 & 0.41    & 0.42     \\ \bottomrule
\end{tabular}
\caption{Popularity prediction performance across all 24 experimental conditions on the held-out test set. Loss strategies are equal-weight sum (Equal), manually weighted (Weighted), and uncertainty-based learned weighting~\cite{kendall2018multi} (Uncert.). FC refers to the number of shared fully connected layers. Mode is either segment or song level, and Task is either popularity-only or full (all seven branches).}
\label{table1}
\end{table*}

%% file: tables/aesthetics.tex
\begin{table}[htb!]
\footnotesize
\centering
\begin{tabular}{l|rrrr}
\toprule
 & MSE & MAE & Pearson & Spearman \\
\midrule
\textbf{Model A} & & & & \\
Coherence & 0.206 & 0.338 & 0.665 & 0.687 \\
Musicality & 0.220 & 0.358 & 0.674 & 0.689 \\
Memorability & 0.250 & 0.377 & 0.670 & 0.690 \\
Clarity & 0.224 & 0.359 & 0.678 & 0.696 \\
Naturalness & 0.206 & 0.344 & 0.696 & 0.712 \\
\midrule
\textbf{Model B} & & & & \\
Coherence & 0.222 & 0.355 & 0.617 & 0.640 \\
Musicality & 0.240 & 0.379 & 0.619 & 0.636 \\
Memorability & 0.271 & 0.398 & 0.618 & 0.639 \\
Clarity & 0.241 & 0.377 & 0.629 & 0.647 \\
Naturalness & 0.225 & 0.365 & 0.637 & 0.656 \\
\midrule
\textbf{Model C} & & & & \\
Coherence & \textbf{0.166} & \textbf{0.304} & \textbf{0.734} & \textbf{0.754} \\
Musicality & \textbf{0.178} & \textbf{0.323} & \textbf{0.739} & \textbf{0.752} \\
Memorability & \textbf{0.203} & \textbf{0.341} & \textbf{0.735} & \textbf{0.751} \\
Clarity & \textbf{0.179} & \textbf{0.322} & \textbf{0.745} & \textbf{0.760} \\
Naturalness & \textbf{0.167} & \textbf{0.312} & \textbf{0.751} & \textbf{0.765} \\
\midrule
\textbf{Model D} & & & & \\
Coherence & 0.219 & 0.347 & 0.643 & 0.669 \\
Musicality & 0.236 & 0.371 & 0.641 & 0.659 \\
Memorability & 0.265 & 0.388 & 0.643 & 0.666 \\
Clarity & 0.235 & 0.368 & 0.650 & 0.672 \\
Naturalness & 0.221 & 0.357 & 0.660 & 0.680 \\
\midrule
\textbf{Model E} & & & & \\
Coherence & 0.236 & 0.362 & 0.592 & 0.623 \\
Musicality & 0.260 & 0.391 & 0.588 & 0.609 \\
Memorability & 0.289 & 0.408 & 0.590 & 0.616 \\
Clarity & 0.257 & 0.386 & 0.601 & 0.626 \\
Naturalness & 0.240 & 0.374 & 0.614 & 0.637 \\
\midrule
\textbf{Model F} & & & & \\
Coherence & 0.198 & 0.327 & 0.687 & 0.714 \\
Musicality & 0.212 & 0.349 & 0.688 & 0.706 \\
Memorability & 0.239 & 0.366 & 0.691 & 0.712 \\
Clarity & 0.211 & 0.345 & 0.699 & 0.719 \\
Naturalness & 0.199 & 0.337 & 0.705 & 0.725 \\
\bottomrule
\end{tabular}
\caption{Song-level aesthetic evaluation performance across models. Lower MSE and MAE indicate better prediction accuracy, while higher Pearson and Spearman correlations indicate stronger alignment.}
\label{tab:song_level_metrics}
\end{table}

%% file: tables/arena.tex
\begin{table}[h!]
\footnotesize
\centering
\begin{tabular}{l|rrr|rrr}
\toprule
 &  & \textbf{AUC} & &  & \textbf{F1} &  \\
\textbf{Model} & \textbf{Overall} & \textbf{Instr}. & \textbf{Vocal} & \textbf{Overall} & \textbf{Instr}. & \textbf{Vocal} \\
\midrule
\multicolumn{5}{l}{\textit{Naive rules}}  & &\\
Likes  & 0.518 & 0.500 & \textbf{0.562} & 0.476 & 0.454 & 0.513 \\
Streams& \textbf{0.557} & \textbf{0.598} & 0.482 & \textbf{0.509} & \textbf{0.533} & 0.470 \\
Aesth. & 0.527 & 0.531 & 0.519 & 0.498 & 0.489 & 0.513 \\
All  & 0.535 & 0.540 & 0.536 & 0.499 & 0.484 & \textbf{0.524} \\
\midrule
\multicolumn{5}{l}{\textit{Model C -- without aesthetics features}}  & &\\
LR & 0.600 & 0.624 & 0.562 & 0.544 & 0.542 & \textbf{0.546} \\
RF & 0.558 & 0.613 & 0.391 & \textbf{0.549} & \textbf{0.579} & 0.510 \\
XGB & 0.562 & 0.606 & 0.473 & 0.513 & 0.536 & 0.479 \\
AdaB & 0.540 & 0.584 & 0.432 & 0.255 & 0.366 & 0.017 \\
SVM & \textbf{0.614} &\textbf{ 0.638} & \textbf{0.572} & 0.524 & 0.532 & 0.511 \\
\midrule
\multicolumn{5}{l}{\textit{Model C -- with aesthetics features}}  & &\\
LR & 0.613 & 0.644 & 0.556 & 0.555 & 0.564 & 0.540 \\
RF & 0.592 & 0.642 & 0.462 & 0.521 & 0.570 & 0.441 \\
XGB & 0.563 & 0.595 & 0.493 & 0.513 & 0.536 & 0.477 \\
AdaB & 0.574 & 0.627 & 0.413 & 0.400 & 0.538 & 0.017 \\
SVM & \textbf{0.642} & \textbf{0.686} & \textbf{0.560} & \textbf{0.595} &\textbf{ 0.625} & \textbf{0.544} \\
\bottomrule
\end{tabular}
\caption{10-fold cross validation results for predicting pairwise preferences on the Music Arena dataset. LR stands for Logistic Regression, RF for Random Forest, XGB for XGBoost, and AdaB for AdaBoost. 
}
\label{tab:arena}
\end{table}



%% file: main.bbl
\begin{thebibliography}{10}
\providecommand{\url}[1]{#1}
\csname url@samestyle\endcsname
\providecommand{\newblock}{\relax}
\providecommand{\bibinfo}[2]{#2}
\providecommand{\BIBentrySTDinterwordspacing}{\spaceskip=0pt\relax}
\providecommand{\BIBentryALTinterwordstretchfactor}{4}
\providecommand{\BIBentryALTinterwordspacing}{\spaceskip=\fontdimen2\font plus
\BIBentryALTinterwordstretchfactor\fontdimen3\font minus \fontdimen4\font\relax}
\providecommand{\BIBforeignlanguage}[2]{{%
\expandafter\ifx\csname l@#1\endcsname\relax
\typeout{** WARNING: IEEEtran.bst: No hyphenation pattern has been}%
\typeout{** loaded for the language `#1'. Using the pattern for}%
\typeout{** the default language instead.}%
\else
\language=\csname l@#1\endcsname
\fi
#2}}
\providecommand{\BIBdecl}{\relax}
\BIBdecl

\bibitem{herremans2020hit}
D.~Herremans and T.~Bergmans, ``Hit song prediction based on early adopter data and audio features,'' \emph{arXiv preprint arXiv:2010.09489}, 2020.

\bibitem{yao2025songeval}
J.~Yao, G.~Ma, H.~Xue, H.~Chen, C.~Hao, Y.~Jiang, H.~Liu, R.~Yuan, J.~Xu, W.~Xue \emph{et~al.}, ``Songeval: A benchmark dataset for song aesthetics evaluation,'' \emph{arXiv preprint arXiv:2505.10793}, 2025.

\bibitem{tjandra2025meta}
A.~Tjandra, Y.-C. Wu, B.~Guo, J.~Hoffman, B.~Ellis, A.~Vyas, B.~Shi, S.~Chen, M.~Le, N.~Zacharov \emph{et~al.}, ``Meta audiobox aesthetics: Unified automatic quality assessment for speech, music, and sound,'' \emph{arXiv preprint arXiv:2502.05139}, 2025.

\bibitem{li2024mert}
\BIBentryALTinterwordspacing
Y.~LI, R.~Yuan, G.~Zhang, Y.~Ma, X.~Chen, H.~Yin, C.~Xiao, C.~Lin, A.~Ragni, E.~Benetos, N.~Gyenge, R.~Dannenberg, R.~Liu, W.~Chen, G.~Xia, Y.~Shi, W.~Huang, Z.~Wang, Y.~Guo, and J.~Fu, ``{MERT}: Acoustic music understanding model with large-scale self-supervised training,'' in \emph{The Twelfth International Conference on Learning Representations}, 2024. [Online]. Available: \url{https://openreview.net/forum?id=w3YZ9MSlBu}
\BIBentrySTDinterwordspacing

\bibitem{udio2026}
{Udio, Inc.}, ``Udio,'' \url{https://www.udio.com}, 2026, online; accessed 22 April 2026.

\bibitem{suno2026}
{Suno, Inc.}, ``Suno,'' \url{https://suno.com}, 2026, online; accessed 22 April 2026.

\bibitem{kimmusic}
Y.~Kim, W.~Chi, A.~N. Angelopoulos, W.-L. Chiang, K.~Saito, S.~Watanabe, Y.~Mitsufuji, and C.~Donahue, ``Music arena: Live evaluation for text-to-music,'' in \emph{The Thirty-ninth Annual Conference on Neural Information Processing Systems Creative AI Track: Humanity}, 2025.

\bibitem{pachet2008hit}
F.~Pachet and P.~Roy, ``Hit song science is not yet a science.'' in \emph{ISMIR}, 2008, pp. 355--360.

\bibitem{herremans2014dance}
D.~Herremans, D.~Martens, and K.~S{\"o}rensen, ``Dance hit song prediction,'' \emph{Journal of New Music Research, Special Issue on Music and Machine Learning}, vol.~43, no.~3, pp. 291--302, 2014.

\bibitem{yang2017revisiting}
L.~C. Yang, S.~Y. Chou, J.~Y. Liu, Y.~H. Yang, and Y.~A. Chen, ``Revisiting the problem of audio-based hit song prediction using convolutional neural networks,'' in \emph{Proc. IEEE Int. Conf. Acoust., Speech, Signal Process. (ICASSP)}, New Orleans, LA, USA, 2017, pp. 621--625.

\bibitem{middlebrook2019song}
K.~Middlebrook and K.~Sheik, ``Song hit prediction: Predicting billboard hits using spotify data,'' \emph{arXiv preprint arXiv:1908.08609}, 2019.

\bibitem{kim2021music}
J.~Kim, ``Music popularity prediction through data analysis of music’s characteristics,'' \emph{International Journal of Science, Technology and Society}, vol.~9, no.~5, pp. 239--244, 2021.

\bibitem{sebastian2024beyond}
N.~Sebastian, F.~Mayer \emph{et~al.}, ``Beyond beats: A recipe to song popularity? a machine learning approach,'' \emph{arXiv preprint arXiv:2403.12079}, 2024.

\bibitem{martin2020multimodal}
D.~Mart{\'i}n-Guti{\'e}rrez, G.~H. Pe{\~n}aloza, A.~Belmonte-Hern{\'a}ndez, and F.~{\'A}. Garc{\'i}a, ``A multimodal end-to-end deep learning architecture for music popularity prediction,'' \emph{IEEE Access}, vol.~8, pp. 39\,361--39\,374, 2020.

\bibitem{zhao2023analysis}
M.~Zhao, M.~Harvey, D.~Cameron, F.~Hopfgartner, and V.~J. Gillet, ``An analysis of classification approaches for hit song prediction using engineered metadata features with lyrics and audio features,'' in \emph{International Conference on Information}.\hskip 1em plus 0.5em minus 0.4em\relax Springer, 2023, pp. 303--311.

\bibitem{cabansag2025prediction}
I.~J. Cabansag and P.~Ntegeka, ``Prediction of spotify chart success using audio and streaming features,'' \emph{arXiv preprint arXiv:2508.11632}, 2025.

\bibitem{yee2022predicting}
Y.~K. Yee and M.~Raheem, ``Predicting music popularity using spotify and youtube features,'' \emph{Indian Journal of Science and Technology}, vol.~15, no.~36, pp. 1786--1799, 2022.

\bibitem{kim2014nowplaying}
Y.~Kim, B.~Suh, and K.~Lee, ``\# nowplaying the future billboard: Mining music listening behaviors of twitter users for hit song prediction,'' in \emph{Proceedings of the first international workshop on Social media retrieval and analysis}, 2014, pp. 51--56.

\bibitem{tsiara2020using}
E.~Tsiara and C.~Tjortjis, ``Using twitter to predict chart position for songs,'' in \emph{IFIP international conference on artificial intelligence applications and innovations}.\hskip 1em plus 0.5em minus 0.4em\relax Springer, 2020, pp. 62--72.

\bibitem{aum2023can}
J.~Aum, J.~Kim, and E.~Park, ``Can we predict the billboard music chart winner? machine learning prediction based on twitter artist-fan interactions,'' \emph{Behaviour \& Information Technology}, vol.~42, no.~6, pp. 775--788, 2023.

\bibitem{rompolas2024predicting}
G.~Rompolas, A.~Smpoukis, E.~Kafeza, and C.~Makris, ``Predicting song popularity through machine learning and sentiment analysis on social networks,'' in \emph{IFIP International Conference on Artificial Intelligence Applications and Innovations}.\hskip 1em plus 0.5em minus 0.4em\relax Springer, 2024, pp. 314--324.

\bibitem{wu2024leveraging}
Y.~Wu, ``Leveraging artificial intelligence for predicting music popularity using social media.'' \emph{El Profesional de la Informaci{\'o}n}, vol.~33, no.~5, 2024.

\bibitem{singhi2015can}
A.~Singhi and D.~G. Brown, ``Can song lyrics predict hits,'' in \emph{Proceedings of the 11th International Symposium on Computer Music Multidisciplinary Research}, 2015, pp. 457--471.

\bibitem{dhanaraj2005automatic}
R.~Dhanaraj and B.~Logan, ``Automatic prediction of hit songs.'' in \emph{Ismir}, 2005, pp. 488--491.

\bibitem{choudhary2025lyrics}
Y.~Choudhary, P.~Rao, and P.~Bhattacharyya, ``Lyrics matter: Exploiting the power of learnt representations for music popularity prediction,'' \emph{arXiv preprint arXiv:2512.05508}, 2025.

\bibitem{vavaroutsos2023hsp}
P.~Vavaroutsos and P.~Vikatos, ``Hsp-tl: a deep metric learning model with triplet loss for hit song prediction,'' in \emph{2023 31st European Signal Processing Conference (EUSIPCO)}.\hskip 1em plus 0.5em minus 0.4em\relax IEEE, 2023, pp. 146--150.

\bibitem{reisz2024quantifying}
N.~Reisz, V.~D. Servedio, and S.~Thurner, ``Quantifying the impact of homophily and influencer networks on song popularity prediction,'' \emph{Scientific Reports}, vol.~14, no.~1, p. 8929, 2024.

\bibitem{li2021lstm}
K.~Li, M.~Li, Y.~Li, and M.~Lin, ``Lstm-rpa: A simple but effective long sequence prediction algorithm for music popularity prediction,'' \emph{arXiv preprint arXiv:2110.15790}, 2021.

\bibitem{liu2022music}
X.~Liu, ``Music trend prediction based on improved lstm and random forest algorithm,'' \emph{Journal of Sensors}, vol. 2022, no.~1, p. 6450469, 2022.

\bibitem{merritt2023accurately}
S.~H. Merritt, K.~Gaffuri, and P.~J. Zak, ``Accurately predicting hit songs using neurophysiology and machine learning,'' \emph{Frontiers in artificial intelligence}, vol.~6, p. 1154663, 2023.

\bibitem{arora2024soundtrack}
S.~Arora and R.~Rani, ``Soundtrack success: Unveiling song popularity patterns using machine learning implementation,'' \emph{SN Computer Science}, vol.~5, no.~3, p. 278, 2024.

\bibitem{xiong2023comprehensive}
Z.~Xiong, W.~Wang, J.~Yu, Y.~Lin, and Z.~Wang, ``A comprehensive survey for evaluation methodologies of ai-generated music,'' \emph{arXiv preprint arXiv:2308.13736}, 2023.

\bibitem{kader2025survey}
F.~B. Kader and S.~Karmaker, ``A survey on evaluation metrics for music generation,'' \emph{arXiv preprint arXiv:2509.00051}, 2025.

\bibitem{kilgour2018fr}
K.~Kilgour, M.~Zuluaga, D.~Roblek, and M.~Sharifi, ``Fr$\backslash$'echet audio distance: A metric for evaluating music enhancement algorithms,'' \emph{arXiv preprint arXiv:1812.08466}, 2018.

\bibitem{zhu2026muq}
D.~Zhu and Z.~Li, ``Muq-eval: An open-source per-sample quality metric for ai music generation evaluation,'' \emph{arXiv preprint arXiv:2603.22677}, 2026.

\bibitem{kendall2018multi}
A.~Kendall, Y.~Gal, and R.~Cipolla, ``Multi-task learning using uncertainty to weigh losses for scene geometry and semantics,'' in \emph{Proceedings of the IEEE conference on computer vision and pattern recognition}, 2018, pp. 7482--7491.

\bibitem{papaioannou2025universal}
C.~Papaioannou, E.~Benetos, and A.~Potamianos, ``Universal music representations? evaluating foundation models on world music corpora,'' \emph{arXiv preprint arXiv:2506.17055}, 2025.

\bibitem{sonauto2026}
{Sonauto}, ``Sonauto,'' \url{https://sonauto.ai}, 2026, online; accessed 22 April 2026.

\bibitem{gong2025ace}
J.~Gong, S.~Zhao, S.~Wang, S.~Xu, and J.~Guo, ``Ace-step: A step towards music generation foundation model,'' \emph{arXiv preprint arXiv:2506.00045}, 2025.

\bibitem{elevenlabs2026}
ElevenLabs, ``Elevenlabs,'' \url{https://elevenlabs.io/}, 2026, online; accessed 22 April 2026.

\bibitem{copet2023musicgen}
J.~Copet, F.~Kreuk, I.~Gat, T.~Remez, D.~Kant, G.~Synnaeve, Y.~Adi, and A.~D{\'e}fossez, ``Simple and controllable music generation,'' in \emph{Advances in Neural Information Processing Systems}, 2023.

\bibitem{riffusion2025fuzz}
\BIBentryALTinterwordspacing
R.~Team, ``Riffusion fuzz: State-of-the-art diffusion transformer for creating and editing music,'' 2025. [Online]. Available: \url{https://riffusion.com}
\BIBentrySTDinterwordspacing

\bibitem{deepmind2025lyria}
\BIBentryALTinterwordspacing
G.~DeepMind, ``Lyria realtime,'' 2025. [Online]. Available: \url{https://magenta.withgoogle.com/lyria-realtime}
\BIBentrySTDinterwordspacing

\end{thebibliography}
